\documentclass[useAMS,usenatbib,usegraphicx]{mn2e}

\usepackage{graphicx}

\title[Finding red quasars and testing the KX method]{Using a complete spectroscopic survey to find red quasars and test the KX method}
\author[R.J. Jurek, M.J. Drinkwater, P.J. Francis and K.A. Pimbblet]{Russell J. Jurek$^{1}$\thanks{E-mail:
jurek@physics.uq.edu.au}, Michael J. Drinkwater$^{1}$, Paul J. Francis$^{2}$ and Kevin A. Pimbblet$^{1}$.\\
$^{1}$Department of Physics, University of Queensland, Brisbane, QLD 4072, Australia\\
$^{2}$Research School of Astronomy and Astrophysics, Australian National University, Canberra, ACT 0200, Australia}
\begin{document}

\date{Accepted 2006 December 01. Received 2006 December 01; in original form 2006 December 01}

\pagerange{\pageref{firstpage}--\pageref{lastpage}} \pubyear{2006}

\maketitle

\label{firstpage}

\begin{abstract}
We present an investigation of quasar colour-redshift parameter space in order to search for radio-quiet red quasars and to test the ability of a variant of the KX quasar selection method to detect quasars over a full range of colour without bias. This is achieved by combining IRIS2 imaging with the complete Fornax Cluster Spectroscopic Survey to probe parameter space unavailable to other surveys. We construct a new sample of 69 quasars with measured $b_J - K$ colours. We show that the colour distribution of these quasars is significantly different from that of the Large Bright Quasar Survey's quasars at a 99.9\% confidence level. We find 11 of our sample of 69 quasars have signifcantly red colours ($b_J - K \geq 3.5$) and from this, we estimate the red quasar fraction of the $K \leq 18.4$ quasar population to be 31\%, and robustly constrain it to be at least 22\%. We show that the KX method variant used here is more effective than the UVX selection method, and has less colour bias than optical colour-colour selection methods.

\end{abstract}


\section{Introduction}
\label{introduction}
\citet{1995Natur.375..469W} compared the $B - K$ colours of Parkes Half-Jansky Flat-Spectrum Sample (PHJFS) \citep{1997MNRAS.284...85D} quasars to those of Large Bright Quasar Survey (LBQS) \citep{1995AJ....109.1498H} quasars. The PHJFS quasars were found to have a significantly broader and redder distribution of $B - K$ colours than LBQS quasars. This disparity should not exist according to the unified model of Active Galactic Nuclei (AGN) galaxies \citep{1993ARA&A..31..473A}. \citet{1995Natur.375..469W} concluded that there exists a population of red quasars that at that time had not been discovered. \citet{1995Natur.375..469W} investigated this possibility using a simple model of LBQS selection effects, and demonstrated that the LBQS was biased against the detection of red quasars. The bias was caused by the blue magnitude limit of the LBQS, because in a blue-flux limited survey, red quasars need to be intrinsically brighter than blue quasars to be detected \citep{1995Natur.375..469W}. The bias causes fewer red quasar detections, because the luminosity function of quasars is steep at the bright end \citep{1991MNRAS.251..482B}. Other surveys with a blue-flux limit are similarly biassed, independent of the quasar selection method. Using their simple model, \citet{1995Natur.375..469W} estimated that the red quasars missed by previous surveys constitute as much as 80\% of all quasars.

Red quasars are defined by a colour criterion. The best colour to use for this criterion is that made from filters separated by the largest wavelength range possible, because anomalously redder quasar colours stand out more. For the datasets used in this paper, the colour covering the largest wavelength range is $U - K$. We used the next largest colour $b_J - K$, as far more objects, $\sim$ 3 times as many, have a measured $b_J - K$ colour compared to the number with a measured $U - K$ colour. The additional advantage of having used $b_J - K$ is that it is possible to compare the colours of the quasar sample presented here with the $B - K$ colour used in \citet{2004ApJ...607...60G} and the $g - K$ colour of \citet{2004AJ....128.1112H}. Throughout this paper we define a red quasar by the criterion, $b_J - K \geq$ 3.5 \citep{2002ApJ...564..133G}, because when describing the optical-infrared region of quasar spectra with a power-law, $S(\lambda) \propto \lambda^{\alpha}$, this colour corresponds to an effective optical-infrared spectral index of $\alpha \geq 1$. 

The existence of red quasars, other than in radio-selected quasar samples, over most of the $b_J - K$ colour range of the PHJFS quasars, satisfying the above criterion, has already been demonstrated \citep{2004AJ....128.1112H, 2003AJ....126.1131R, 2004ApJ...607...60G}. The \citet{2004ApJ...607...60G} red quasar sample contains the reddest quasars found, with $B - K$ colours of $5 \leq B - K \leq 8$. \citet{2004AJ....128.1112H, 2003AJ....126.1131R} found red quasars in the Sloan Digital Sky Survey (SDSS)\citep{2000AJ....120.1579Y}, and \citet{2004ApJ...607...60G} found them in a combination of the 2-Micron All Sky Survey (2MASS) \citep{2006AJ....131.1163S} and an Automatic Plate Measuring CATalogue (APMCAT) of POSS/UKST sky survey plates \citep{APMCAT}. 

Despite confirming the existence of red quasars, current observations have not explained the mechanism or mechanisms responsible for their existence. Dust obscuration by dust at the quasar redshift, in either the host galaxy or the quasar accretion disk \citep{1995Natur.375..469W}, was the first mechanism proposed for quasar reddening. \citet{1995Natur.375..469W} proposed this mechanism as a correlation between $B - K$ colour and redshift was not observed. The result of reddening a composite optically selected quasar spectral energy distribution (SED) with dust at the quasar redshift, was found to be consistent with the dust mechanism, because the reddening resulted in the SED of a red PHJFS quasar \citep{1995Natur.375..469W}. 

As an alternative, \citet{1998MNRAS.295..451B} suggested that the $B - K$ spread in the \citet{1995Natur.375..469W} sample is partly or entirely caused by host galaxy starlight. Quasars with flat-radio-spectra are typically hosted by later galaxy types \citep{1998MNRAS.295..451B}, and the old stellar population of late type galaxies is a stronger emitter in K than B, reddening quasars \citep{1998MNRAS.295..451B}. An analysis in \citet{1998MNRAS.301..975M} of the host galaxy contribution to the red $B - K$ colours of the PHJFS quasars in \citet{1995Natur.375..469W} was inconclusive, but suggested host galaxy contribution was insufficient to cause the red $B - K$ colours observed. Further work in \citet{2006MNRAS.367..717M} confirmed that host galaxy starlight can definitely redden quasars, particularly low luminosity and low redshift resolved quasars. In accordance with \citet{2006MNRAS.367..717M}, in this paper we use the criteria that luminous unresolved quasars, which are not located at low redshift, are unlikely to be reddened by host galaxy contribution.

Alternatively, \citet{1996Natur.379..304S} proposed that a synchrotron component of quasar emission created a K-band excess, reddening quasars. An analysis of 157 PHJFS quasars and 12 LBQS quasars found the shape of the SEDs was consistent with both line-of-sight dust and synchrotron emission mechanisms \citep{2000PASA...17...56F}. In \citet{2001AJ....121.2843B} it was suggested that because any synchrotron component to quasar emission is weaker in radio-quiet quasars than in radio-loud quasars, any synchrotron emission will only contribute to a small fraction of any reddening. Based on the quasar sample contained therein, \citet{2001AJ....121.2843B} suggested that any radio-quiet quasar redder than $b_J - K \geq$ 3.7 is too red to be the result of synchrotron emission, and is most likely the result of dust obscuration. Throughout this paper we use the criterion that any red quasar that is also radio-quiet, is unlikely to be reddened by synchrotron emission.

Finally, \citet{1997MNRAS.288..138M} suggested that a small subset of red quasars was caused by the lensing of a quasar by a dusty galaxy. Not only does the dust in the lens redden the quasar but the dusty galaxy potentially magnifies the host galaxy starlight contribution \citep{2002ApJ...564..133G}. This last mechanism was proposed after \citet{1997MNRAS.288..138M} observed a similar phenomenon to \citet{1995Natur.375..469W}; with radio-selected lenses having redder colours than optically selected lenses. 

Whichever mechanism creates red quasars, they are possibly an evolutionary stage. After observing low-ionisation Broad Absorption Line (BAL) quasars, \citet{2002AJ....123.2925L} suggested a model of quasar evolution where a quasar occupies a thick dusty torus after forming, eventually the torus dissipates leaving behind an optically blue quasar. Alternately, quasars may emerge from dusty starburst galaxies, residing in the dust stirred up by mergers \citep{2005AAS...20717401U}. Either evolutionary sequence would naturally create a situation where quasars are reddened by dust at the quasar redshift. A third evolutionary path has been proposed where red quasars are an evolutionary link between Ultra-Luminous Infrared Galaxies (ULIRGs) and optically selected quasars \citep{2006Ap&SS.302...17C}. Determining the fraction of red quasars and any dependence on redshift, luminosity, radio flux, etc. will indicate if red quasars are an evolutionary stage, and which evolutionary path red quasars belong to.

Depending on the mechanism, red quasars might explain the quasar contribution to the X-Ray Background (XRB). Using a theoretical model, \citet{1994MNRAS.270L..17M} showed that a ratio of two to three times as many dust obscured to unobscured active galactic nuclei (AGN) galaxies can adequately account for the existence of the XRB. The model also accounts for the observed spectrum and the source counts in the soft and hard X-ray bands. The latest theoretical models of the XRB require a smaller ratio, lying somewhere between 0.6 to 1.5 \citep{2007A&A...463...79G}. If it can be shown that red quasars are the result of dust obscuration, and that the red quasar fraction is within 37.5\% to 60\%, then red quasars are the most probable source of the quasar contribution to the XRB.

The existence of red quasars has various implications, the main one is that it can no longer be assumed that existing quasar samples are representative of the entire quasar population. The statistics of a sample reflects the properties of the sub-set of a population from which the sample is drawn, previous quasar samples, such as the LBQS, are biassed to the inclusion of blue quasars; therefore, the statistics of such samples predominantly reflects the blue quasar population. Use of existing quasar datasets requiring a sample that is representative of the entire quasar population is therefore subject to review. Direct examples of this are that the luminosity function of quasars and the change in number density with redshift need to be re-evaluated while including red quasars.

In order to learn more about red quasars, we need to spectroscopically observe quasars over their entire colour range. To achieve this, future surveys will have to select quasar candidates for spectroscopic follow-up using a technique that is not biased against either red or blue quasars. One candidate for a suitable selection technique is the KX method, developed in \citet{2000MNRAS.312..827W}. The KX method utilises the power-law nature of quasar SEDs at long wavelengths combined with morphological classification. Stars experience a turnover in their SED in H-band while quasars follow a power-law, allowing discrimination between stars and quasars. The stellar morphology of quasars discriminates between quasars and galaxies. Observationally, this requires constructing a colour-colour plot of all objects with a stellar morphology, using an optical colour and a colour that straddles the H-band. 

In this paper we explore regions of the quasar colour-redshift space unavailable in previous work, and test a KX method variant. We aim to find the boundaries of the quasar colour-redshift parameter space that is occupied. In doing so, we will improve the estimate of the red quasar fraction and possibly shed light on the mechanism that creates red quasars. In testing a KX method variant, we intend to verify if it is a viable method of selecting quasars, without bias, from the entire quasar colour-redshift parameter space. We intend to apply a KX method variant to our quasar sample, and then contrast it with established methods of selecting potential quasars for spectroscopic follow-up. 

To achieve these goals we use a sub-sample of the quasars identified in the Fornax Cluster Spectroscopic Survey (FCSS) \citep{2000A&A...355..900D}, a spectroscopic survey of all extended objects to $b_J = 19.8$ and point sources to $b_J = 21.5$. The sub-sample consists of all quasars that match to an object in Ks-band imaging to a depth of K = 18.4, obtained using the Infrared Imager and Spectrograph 2 (IRIS2) instrument on the 3.92-m Anglo-Australian Telescope (AAT). The Ks-band imaging is deeper than 2MASS allowing us to explore parts of the $b_J - K$ colour space unavailable in \citet{2004ApJ...607...60G}, which was limited to $b_J - K \geq$ 5 by the K = 14.5 magnitude limit of 2MASS. Furthermore, the quasar sample used covers a larger range of redshift than was examined in \citet{2004AJ....128.1112H,2003AJ....126.1131R}, which were limited to z $\leq$ 2.2. Because the FCSS used no selection criteria, our quasar sample is bias free, and ideal for testing the ability of the KX method to select quasars from a K magnitude limited survey. Additionally, most of the FCSS objects have a U magnitude as well as a $b_J$ magnitude, allowing the KX method to be directly contrasted with traditional methods for optically selecting quasars. 

The structure of this paper is as follows. In Section \ref{data} the datasets used in this paper are described. Section \ref{redquasars} details the detection of red quasars in the quasar sample and an analysis of the quasar sample. This analysis involves a comparison of the $b_J - K$ colours of the sample quasars to LBQS quasars and a determination of the effect of our selection on the quasar sample constructed here. Section \ref{redquasars} concludes with the calculation of the fraction of red quasars. The effectiveness of the KX method is assessed in Section \ref{testKX} and following this are our conclusions in Section \ref{conclusion}.

\section[]{Photometric and Spectroscopic Data}
\label{data}
The sample of quasars used here and their photometric information were obtained by combining three datasets. Astrometry and optical photometry were obtained from the Automatic Plate Measuring CATalogue (APMCAT) of POSS/UKST sky survey plates \citep{APMCAT}. IRIS2 imaging was used to obtain the K-band photometry used here. The necessary spectroscopic identifications were provided by the Fornax Cluster Spectroscopic Survey (FCSS) catalogue of \citet{2000A&A...355..900D}. The datasets were combined by independently matching the astrometry of K-band catalogue objects and FCSS objects to that of APMCAT catalogue objects. In the rest of this section, we describe the datasets in more detail.

The FCSS catalogue is a blind spectroscopic survey of Fornax objects in the APMCAT catalogue. The FCSS is an ideal source of spectroscopic identifications because the lack of target selection precludes biasing. In the FCSS, spectroscopy was carried out on all extended sources to $b_J \leq 19.8$ and all point sources to $b_J \leq 21.5$. Not every source was spectroscopically observed, and the fraction of observed sources is referred to as the spectroscopic completeness. The FCSS obtained a redshift and spectroscopic identification for more than 90\% of the objects observed, referred to as the redshift completeness. In most cases where a spectroscopic identification could not be obtained, neither was a redshift because of the poor spectrum quality. Because the redshift completeness was consistently high for the entire FCSS, we have combined the redshift completeness with the spectroscopic completeness, and throughout the paper the term `completeness' refers to this combination. The completeness of the FCSS is 96\% for $b_J < 20.5$, 82\% for $20.5 \leq b_J < \leq 21$ and 36\% for  $21 < b_J \leq 21.5$ sources. We account for the completeness by appropriately weighting quasars during our analysis in later sections. 

IRIS2 was used in December 2001 to obtain Ks-band imaging using 60s exposures. The raw data were processed using the Observatory Reduction and Acquisition Control project - Data Reduction pipeline (ORAC-DR) outlined in \citet{1999ASPC..172...11E,1999ASPC..172..171J}. The reduced IRIS2 imaging dataset was then analysed with SExtractor \citep{1996A&AS..117..393B} to obtain a photometric catalogue of corrected isophotal magnitudes. Corrected isophotal magnitudes were used instead of PSF or small aperture magntiudes, because PSF and small aperture magnitudes minimise any possible extended contribution to the flux of sources in the IRIS2 imaging. The Ks-band catalogue magnitude zero-point was then calibrated, by matching point sources in the Ks-band catalogue to the 2MASS point source catalogue. Calibrating also facilitated the conversion from Ks magnitudes to the K-band magnitudes used throughout this paper. After calibration, the magnitude limit of the K-band photometry was found to be 18.4 magnitudes. 

In the FCSS, quasars were spectroscopically identified as objects with broad permitted lines with a measured full width half-maximum $>$ 1100 km $s^{-1}$ \citep{2001MNRAS.324..343M}. After matching the IRIS2 imaging catalogues and the FCSS quasar identifications to the APMCAT object positions using blind matching, to a radius of 10"; a catalogue of all the necessary photometry was created, and all quasars within both the IRIS2 imaging K magnitude limit, and the FCSS spectroscopic identifications $b_J$ magnitude limit were identified. This resulted in a sample of 69 spectroscopically identified quasars with U, $b_J$ and K magnitudes. All 69 of these quasars are APMCAT point sources; therefore, our quasar sample is limited to $K \leq 18.4$ and $b_J \leq 21.5$. Of these 69 quasars, 62 have an R magnitude as required by our KX method variant. The missing R magnitudes are caused by the different magnitude limits in the $b_J$ and R bands of the APMCAT catalogue, 22.5 and 21 magnitudes respectively.

\section{Red quasars}
\label{redquasars}
Combining the APMCAT, IRIS2 imaging and FCSS datasets resulted in a sample of 69 spectroscopically identified quasars. In Section \ref{detection} we examine the $b_J - K$ colours of the quasars to determine which quasars are red. We investigate the uncertainties in the data to ascertain if the uncertainties caused blue quasars to appear red. Section \ref{comparison} compares the colours of the constructed quasar sample to the colours of a sub-set of LBQS quasars to characterise the `reddness' of the quasar sample. Then in Section \ref{selection}, we examine selection effects and estimate the red quasar fraction. Finally, Section \ref{relativebjMk} examines the relative $b_J - k$ colours of our quasar sample, and compares them to the relative colours of the LBQS sub-set.

\subsection{Detection of red quasars}
\label{detection}

\begin{figure}
\includegraphics[width=84mm]{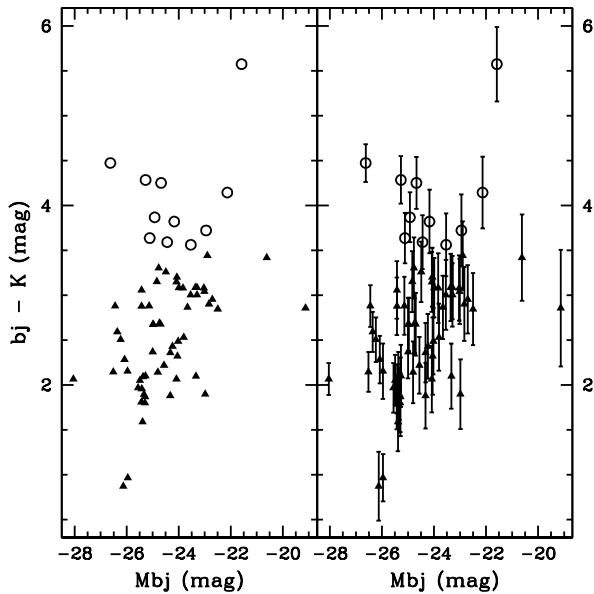}
\caption{A plot of $b_J - K$ colour as a function of absolute $b_J$ magnitude, with (right) and without (left) uncertainties. Quasars with a red, $b_J - K \geq 3.5$, colour are denoted by circles and blue quasars with solid triangles. The $b_J - K$ colour uncertainty is the result of combining in quadrature the photometric uncertainties in the $b_J$ and K mag with the quasar variability $b_J$ magnitude structure function in \citet{1994MNRAS.268..305H}.}
\label{colourvsMag}
\end{figure}

\begin{figure}
\includegraphics[width=84mm]{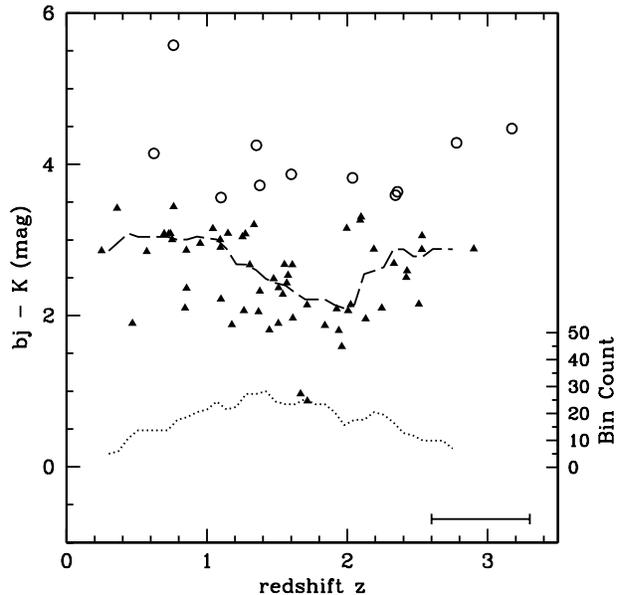}
\caption{bj - K colour vs redshift, z, for the 69 quasars presented here. Blue quasars ($b_J - k < 3.5$) are solid triangles and red quasars ($b_J - K \geq 3.5$) are circles. Using redshift bins of size z $=$ 0.7 (reference image in bottom right corner) centered at points 0.07 apart from z $=$ 0.3 to z $=$ 2.8, the number of quasars and median $b_J - K$ colour is plotted. The median $b_J - K$ colour is plotted as a dashed line and the number of number of quasars in each bin, with a separate axis on the right hand side, is plotted as a dotted line. }
\label{bjMKvsz}
\end{figure}

We show the $b_J - K$ colours of our quasar sample in Figure \ref{colourvsMag}, 11 quasars are red because they satisfy $b_J - K \geq 3.5$, corresponding to $\approx 0.16$ of the 69 quasars with $b_J$ and K magnitudes. These red quasars have colours, 3.5 $\leq b_J - K \leq$ 5.6, which occupy the colour space between \citet{2004ApJ...607...60G} red quasars and LBQS, SDSS red quasars. Using the All-Sky Optical Catalogue of Radio/X-Ray Sources in \citet{2004A&A...427..387F}, we verified the existence (or lack thereof) of a radio-counterpart for all the red quasars discovered here, to a flux limit of 1mJy. We found that 5 of these quasars had no radio-counterpart and the remaining 6 were not in the catalogue, indicating they too are radio-quiet. We concluded that all of the red quasars in Figure \ref{colourvsMag} are radio-quiet quasars. In \citet{2006MNRAS.367..717M} it was demonstrated that the host galaxy contribution to quasar flux is maximal in the K-band for resolved, low-luminosity quasars at $z<1$. $b_J - K$ is plotted against redshift in Figure \ref{bjMKvsz}; and a dotted line tracks the number of quasars as a function of redshift, peaking at z $\approx 1.5$. The dashed line in Figure \ref{bjMKvsz} traces the median quasar colour with redshift, demonstrating a dependence of the observed $b_J - K$ colour on redshift. Section \ref{relativebjMk} examines the dependence of the observed $b_J - K$ colour on redshift. As the red quasars found in Figure \ref{colourvsMag} are all luminous unresolved APMCAT sources, and the majority are luminous quasars with $z > 1$ in Figure \ref{bjMKvsz}; any host galaxy contribution to the $b_J - K$ colours of the red quasars will be minimal. If a significant host galaxy contribution was present in the red quasars identified in Figure \ref{colourvsMag}, then the $b_J - K$ colours of the red quasars would become increasingly redder as both luminosity and redshift decreased. In Figures \ref{colourvsMag} and \ref{bjMKvsz}, a clear trend of increasingly red $b_J - K$ colours with decreasing luminosity and redshift is not apparent. The 11 red radio-quiet quasars detected here are most likely the result of reddening by dust located at the quasar redshift.
 
We investigated the uncertainties in the observed $b_J - K$ colours, to determine if the uncertainties caused blue quasars to appear red. Uncertainty in the $b_J - K$ colours in Figure \ref{colourvsMag} has two sources, photometric uncertainties in the $b_J$ and K magnitudes and the effect of quasar variability on multi-epoch observations. The photometric uncertainty in the $b_J$ magnitudes was obtained from \citet{2000A&A...355..900D} where it was found to be 0.1 mag. For the K-band magnitudes the uncertainty was calculated by SExtractor during the data reduction. To account for the effect of quasar variability we estimated the uncertainty in the $b_J$ photometry, due to it being observed at an earlier epoch to the K band photometry, using the structure function of \citet{1994MNRAS.268..305H}, 
\begin{equation}
|\Delta m_{b_J}| = [0.155 + 0.023 (M_B + 25.7)] \Delta t_{obs}^{0.18} (1 + z)
\end{equation}
where $\Delta t_{obs}$ is the time difference between the multi-epoch observations in the observed frame. In doing this we effectively move the $b_J$ photometry to the epoch of the K photometry. To evaluate the structure function we used $\Delta t_{obs} =$ 25 years and calculated the absolute $b_J$ magnitudes using $k(z) = -1.25log(1+z)$ \citep{2001MNRAS.324..343M}, $\Omega_m = 0.3$, $\Omega_{\Lambda} = 0.7$ and $H_0 = 72$ km $s^{-1}$ $Mpc^{-1}$. The $b_J - K$ colour uncertainties in Figure \ref{colourvsMag} are the combination of the photometric uncertainties and the structure function result in quadrature. Based on Figure \ref{colourvsMag}, some of the red quasars found here are potentially blue quasars, that appeared red because of uncertainties in their $b_J - K$ colours.

\begin{table}
\caption{Completeness limits and associated weighting applied to a quasar as a function of $b_J$ magnitude.}
\label{weightingtable}
\begin{tabular}{lcc}
\hline
$b_J$ magnitude & completeness & weight applied to quasar\\
\hline
$b_J < 20.5$ & 96\% & 1.04\\
$20.5 \leq b_J \leq 21$ & 82\% & 1.22\\
$21 < b_J \leq 21.5$ & 36\% & 2.78\\
\hline
\end{tabular}
\end{table}

\begin{figure}
\includegraphics[width=84mm]{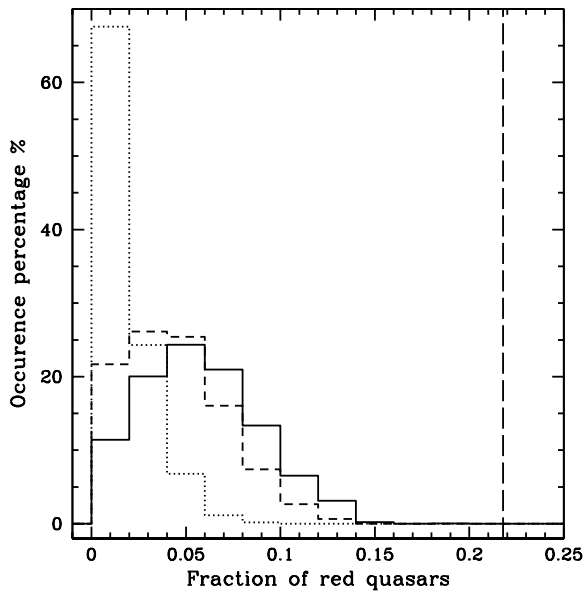}
\caption{Results of scattering the $b_J - K$ colours of 51 blue ($b_J - K <$ 3.5) quasars with Gaussians scaled by the photometric uncertainties (dotted line), quasar variability structure function (short dashed line) and both combined in quadrature (solid line). Plotted are histograms of the percentage of 10,000 iterations that resulted in a given fraction of the blue quasars being scattered so as to have red ($ b_J - K \geq$ 3.5) colours. The long dashed line marks the fraction of red quasars found here.}
\label{TestRedAll}
\end{figure}

We explicitly tested if all the red quasars observed here are the result of photometric uncertainties and/or quasar variability by scattering the measured $b_J - K$ colours of the 51 blue quasars ($ b_J - K <$ 3.5) by a Gaussian, scaled so that 1-sigma equals the $b_J - K$ uncertainty, and calculating the resultant fraction of red quasars; after weighting the quasars by the completeness of the FCSS, using the weights in Table \ref{weightingtable}. The solid line in Figure \ref{TestRedAll} shows the result of 10,000 attempts at scattering the $b_J - K$ colours of the blue quasars; a histogram of the percentage of the 10,000 iterations that resulted in a given red quasar fraction. The red quasar fraction observed here, after weighting the quasars using the weights in Table \ref{weightingtable}, is $\sim$0.22. Based on the percentage of 10,000 iterations that resulted in an equal or greater red quasar fraction, we concluded from Figure \ref{TestRedAll} that we could reject at the 99.9\% confidence level that our observed red quasar fraction is the result of photometric uncertainties and/or quasar variability. 

The solid line in Figure \ref{TestRedAll} has a peak at a non-zero red quasar fraction, suggesting that the $b_J - K$ uncertainties are reddening our quasar sample. To investigate the cause of this potential reddening we again scattered the $b_J - K$ colours of the 51 blue quasars, this time scaling the Gaussians by the $b_J - K$ photometric uncertainties and the effect of quasar variability separately, and the result of 10,000 iterations are shown in Figure \ref{TestRedAll} as a dotted and dashed line respectively. From Figure \ref{TestRedAll} we concluded that quasar variability was potentially reddening our quasar sample as the dashed line in Figure \ref{TestRedAll} peaks at a non-zero red quasar fraction; and that the photometric uncertainties did not redden our quasar sample.

\subsection{Comparison to LBQS quasars}
\label{comparison}
\begin{figure}
\includegraphics[width=84mm]{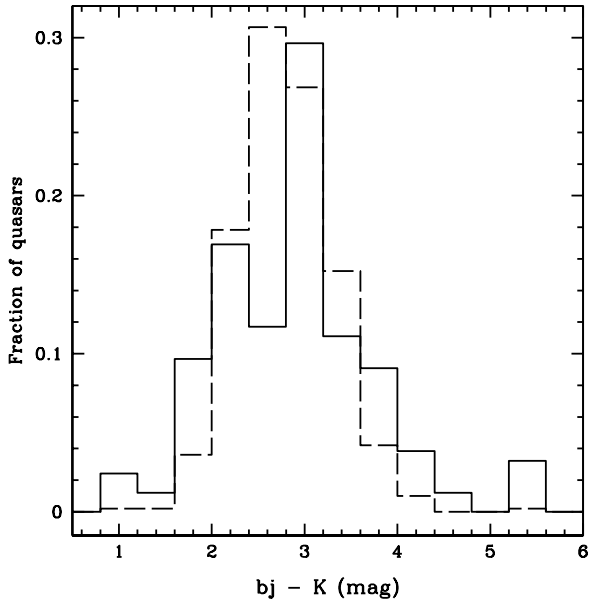}
\caption{Histograms of quasar fraction as a function of $b_J - K$ colour. The solid line is the weighted $b_J - K$ distribution of the 69 quasars in our sample, weighted using the values in Table \ref{weightingtable}. The dashed histogram is the distribution for a sample of 538 LBQS quasars matched to 2MASS. The distribution of the 69 measured $b_J - K$ colours extends noticeably further and redder than the distribution of LBQS $b_J - K$ colours.}
\label{bjMKhistogram}
\end{figure}

\begin{figure}
\includegraphics[width=84mm]{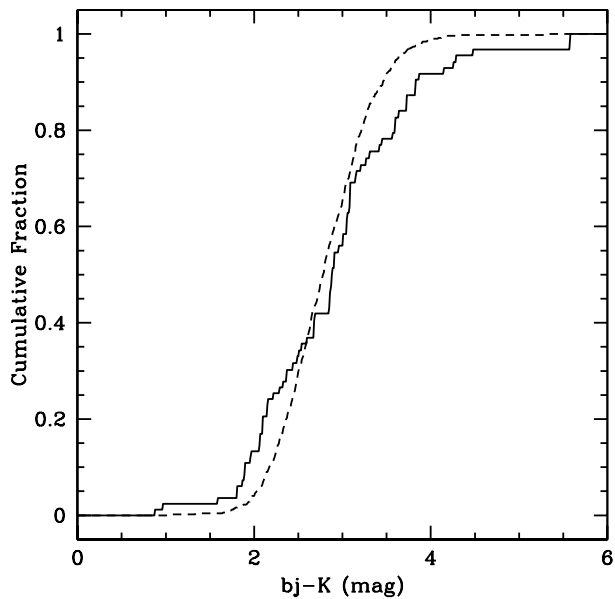}
\caption{Cumulative distribution function of $b_J - K$ colours. The solid line is for the 69 quasars in our sample, weighted using the values in Table \ref{weightingtable}. The dashed line is for a sample of LBQS quasars matched to 2MASS. A Kuiper test of the two distributions shows them to be different at the 99.9\% confidence level.}
\label{bjMKcdfs}
\end{figure}

The sample of 69 quasars was compared to a sample of 538 LBQS quasars that matched objects in the 2MASS survey. Histograms of both sets of quasar $b_J - K$ colours were plotted in Figure \ref{bjMKhistogram}, where the $b_J - K$ colours for our 69 quasars were weighted by the completeness of the FCSS, using the values shown in Table \ref{weightingtable}. The histograms in Figure \ref{bjMKhistogram} suggest that the quasars presented here have a different $b_J - K$ distribution to the LBQS quasars, covering a larger range and extending noticeably redder. For the weighted $b_J - K$ colours, the average, standard deviation, excess kurtosis and skewness are $\overline{b_J - K} = 2.9$, $\sigma = 0.9$, $-0.48$ and $0.46$ magnitudes respectively. The 538 LBQS quasars have an average, standard deviation, excess kurtosis and skewness of $\overline{b_J - K} = 2.8$, $\sigma = 0.5$, $-2.7$ and $0.04$ magnitudes respectively. Plotting cumulative frequency distributions of the 69 measured $b_J - K$ colours and the 538 LBQS quasars in Figure \ref{bjMKcdfs}, a Kuiper test showed the two distributions to be different at the 99.9\% confidence level. The statistics of the two $b_J - K$ distributions are consistent with our sample having a redder, broader distribution than the LBQS quasars. 

The concern in making this comparison is that any difference between the quasar sample constructed here and the 538 LBQS quasars may be the result of $b_J - K$ uncertainties. To investigate this we scattered the $b_J - K$ colours of the LBQS quasars by the uncertainties in their $b_J - K$ colours, using Gaussians scaled such that 1-$\sigma$ equals the $b_J - K$ uncertainty, and tested if the two distributions were still different using a Kuiper test. The LBQS $b_J - K$ uncertainties were determined by adding in quadrature the $b_J$ magnitude uncertainties in the LBQS, K magnitude uncertainties in 2MASS and the $b_J - K$ uncertainties due to quasar variability. The $b_J - K$ uncertainties due to variability were calculated using the structure function of \citet{1994MNRAS.268..305H}, outlined in Section \ref{detection}, with $\Delta t_{obs} =$ 11 years. Scattering the LBQS $b_J - K$ colours 10,000 times, a Kuiper test showed the distributions of the scattered LBQS $b_J - K$ colours and our quasar samples $b_J - K$ colours were different 99.3\%, 71.7\% and 24.9\% of the time, at the 95\% (2-$\sigma$), 99\% and 99.7\% (3-$\sigma$) confidence levels respectively. Based on the results of scattering the LBQS $b_J - K$ colours, we concluded that $b_J - K$ uncertainties were unlikely to cause the observed difference between our quasar sample and the LBQS quasars.

\subsection{Relative $b_J - K$ colours}
\label{relativebjMk}
\begin{figure}
\includegraphics[width=84mm]{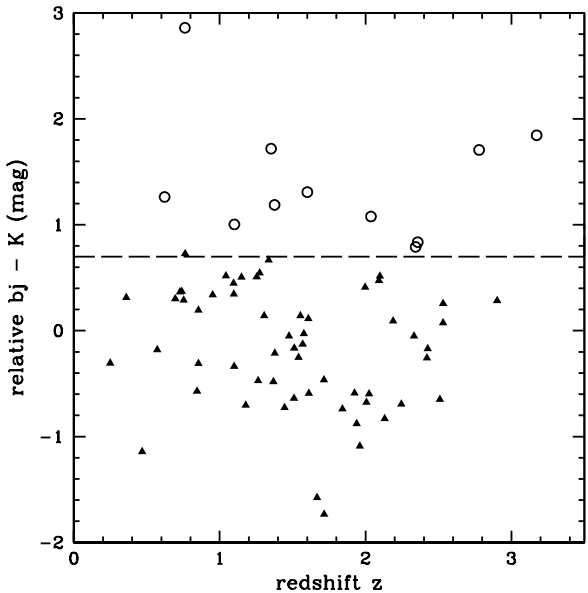}
\caption{Figure \ref{bjMKvsz} re-plotted with the median colour for a given redshift subtratcted to give relative $b_J - K$ colours. Quasars with blue observed $b_J - K$ colours ($b_J - K < 3.5$) are plotted as solid triangles and quasars with red observed $b_J - K$ colours ($b_J - K \geq 3.5$) are plotted as circles. A dashed horizontal line marks the relative $b_J - K$ criterion proposed for red quasars, relative $ b_J - K \geq 0.5$. Four quasars with blue observed $b_J - K$ colours have red relative $b_J - K$ colours, demonstrating the dependence of red/blue quasar classification on redshift when using observed $b_J - K$ colours.}
\label{relbjMKvsz}
\end{figure}

As the observed $b_J - K$ colour is dependent upon redshift, there may be red quasars in our sample that in the observed frame failed to meet the $b_J - K \geq$ 3.5 colour criterion; instead being classified as blue quasars. We investigated this in Figure \ref{relbjMKvsz} by plotting relative $b_J - K$ colours, the quasar $b_J - K$ colours minus the median $b_J - K$ colour for the quasar redshift, essentially applying a first order K-correction. We measured the median $b_J - K$ colour, that was subtracted from both our quasar sample and the LBQS quasar sample, at a given redshift from the LBQS quasar sample; because, in a quasar population where red quasars occupy a tail that extends away from blue quasars, the median $b_J - K$ colour will be dominated by the $b_J - K$ colours of blue quasars. All the quasars in our sample, that are red based on their observed $b_J - K$ ($b_J - K \geq 3.5$) colour have a relative colour such that $b_J - K \geq 0.7$, as such we define this as the relative $b_J - K$ colour criterion for a red quasar. One quasar that is blue, based on its observed $b_J - K$ colour, would have been classified red using its relative $b_J - K$ colour. This quasar demonstrates that observed frame colour criteria are potentially unsuitable for differentiating between blue and red quasars, as commented on in \citet{2006AJ....132.1977H}. It is beyond the capability of this paper to determine whether observed or relative $b_J - K$ colours are better suited to differentiating between blue and red quasars, because of the small size of our quasar sample. 

\begin{figure}
\includegraphics[width=84mm]{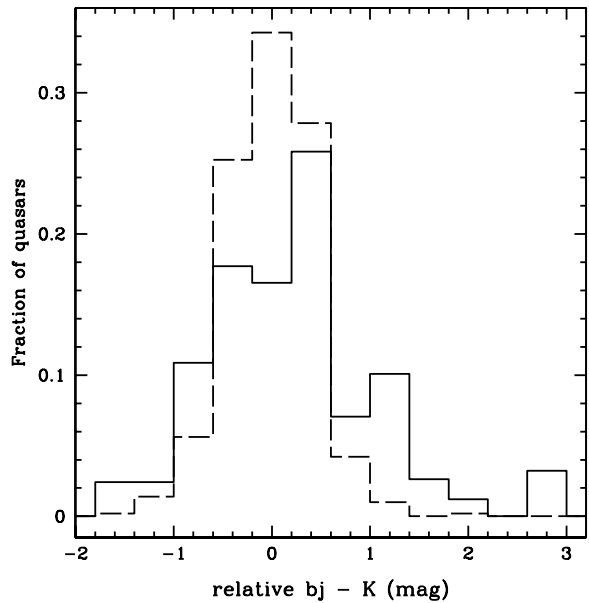}
\caption{Histograms of quasar fraction as a function of relative $b_J - K$ colour. The solid line is the weighted distribution of our samples 69 quasars relative $b_J - K$ colours, weighted using the values in Table \ref{weightingtable}. The dashed histogram is the relative $b_J - K$ distribution for a sample of 538 LBQS quasars matched to 2MASS. The distribution of our quasar samples relative $b_J - K$ colours extends noticeably further and redder than the distribution of LBQS relative $b_J - K$ colours.}
\label{relbjMkHist}
\end{figure}

\begin{figure}
\includegraphics[width=84mm]{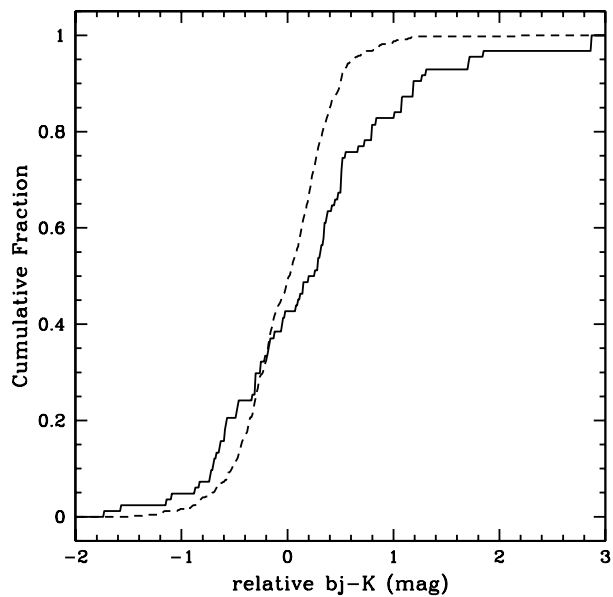}
\caption{Cumulative distribution function of relative $b_J - K$ colours. The solid line is for the relative $b_J - K$ colours of the 69 quasars in our sample, weighted using the values in Table \ref{weightingtable}. The dashed line is for the relative $b_J - K$ colours of a sample of LBQS quasars matched to 2MASS. A Kuiper test of the two distributions shows them to be different at the 99.9\% confidence level.}
\label{relbjMkCDFs}
\end{figure}

To investigate the effect of using relative colours on our comparison, we compared the relative $b_J - K$ colours of our quasar sample to the relative colours of the LBQS quasars. We constructed the sample of relative LBQS colours by subtracting the median LBQS $b_J - K$ colour, at a given redshift, from each LBQS quasar. In Figure \ref{relbjMkHist} we plotted the relative $b_J - K$ distributions of both our sample and the LBQS quasars. In Figure \ref{relbjMkHist} our samples relative $b_J - K$ distribution appears to be broader and redder than the LBQS relative $b_J - K$ distribution; and, compared to Figure \ref{bjMKhistogram}, the small amount of excess blue quasars in our sample compared to the LBQS decreased, while the excess of red quasars compared to the LBQS increased. Our samples relative $b_J - K$ colours have an average, standard deviation, excess kurtosis and skewness of $\overline{b_J - K} = 0.2$ and $\sigma = 0.9$, $-0.6$ and $0.5$ magnitudes respectively. The relative colours of the LBQS quasars have an average, standard deviation, excess kurtosis and skewness of $\overline{b_J - K} = -0.01$, $\sigma = 0.43$, $-2.8$ and $0.01$ magnitudes respectively. Plotting cumulative frequency distributions of the relative $b_J - K$ colours of the 69 quasars in our sample, and the relative $b_J - K$ colours of the LBQS quasars in Figure \ref{relbjMkCDFs}, a Kuiper test showed the two distributions to be different at the 99.9\% confidence level. The statistics of the two relative $b_J - K$ distributions are consistent with our quasar sample having a broader and redder relative $b_J - K$ distribution than the LBQS quasars. Based on the differences between Figures \ref{bjMKhistogram} and \ref{relbjMkHist}, and our statistical analyses here and in Section \ref{comparison}, we determined that whether we used observed or relative quasar colours, the results of comparing our quasar sample to a sample of LBQS quasars is the same, that our sample has a broader, redder $b_J - K$ distribution than the LBQS quasars. The additional information gleaned from our comparison of the relative $b_J - K$ colours is that the small excess of blue quasars in our sample, when compared to the LBQS, appears to be caused by K-correction effects, and is not actually an intrinsic property of the $b_J - K$ distribution. Because the excess of blue quasars in our quasar sample, when compared to the LBQS quasar sample, is probably caused by K-correction effects, the small excess of blue quasars is much less significant than the excess of red quasars when comparing our quasar sample to the LBQS sample.

\subsection{Limits on the red quasar fraction}
\label{selection}
\begin{figure}
\includegraphics[width=84mm]{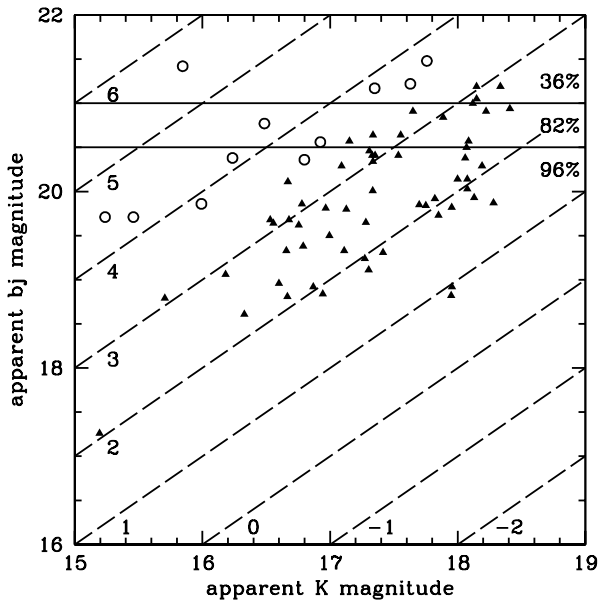}
\caption{bj vs. K for the 69 quasars presented here. Blue quasars (bj - k $< 3.5$) are solid triangles and red quasars (bj - K $\geq$ 3.5) are circles. Lines of constant $b_J - K$ are plotted as dashed lines, indexed as they increase from the bottom right corner. Solid horizontal lines mark the regions of the FCSS completeness as a function of $b_J$ magnitude, and are labeled accordingly.}
\label{bjvsK}
\end{figure}

To understand the effects of our sample selection on the $b_J$ and K magnitude parameter space, we plotted $b_J$ and K magnitudes in Figure \ref{bjvsK}, with completeness limits and lines of constant $b_J - K$ overlaid. The plot in Figure \ref{bjvsK} revealed where in $b_J$ and K magnitude parameter space the number of quasar detections declines; and, by extension revealed where the $b_J - K$ distribution will drop off due to selection effects. The upper left region of Figure \ref{bjvsK} is where we expect the least quasar detections due to the combination of completeness, and the magnitude limits of the $b_J$ and K photometry, referred to collectively as selection effects. A few quasars are found with $b_J - K > 4$ and only one is found satisfying $K < 17$ and $b_J > 20.5$. As the decrease in observed quasar detections coincides with where it is expected to decrease because of selection effects; we conclude from Figure \ref{bjvsK} that the decrease in quasar detections, redder than $ b_J - K = 4$, is a result of sample selection and not an intrinsic property of our quasar sample. This is consistent with the fact that redder quasars than those found here have been observed, such as in \citet{2004ApJ...607...60G}.

\begin{figure}
\includegraphics[width=84mm]{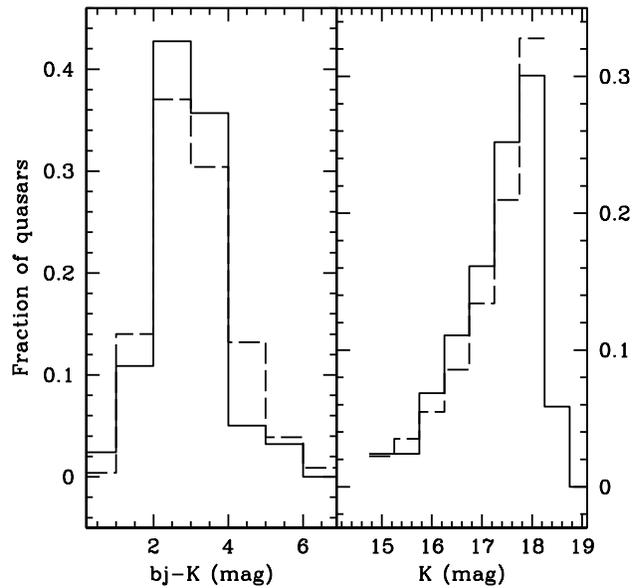}
\caption{Plot of the observed $b_J - K$ (left) and K (right) distributions of our quasar sample, and the distributions predicted by our model of quasar density in the $b_J - K$ vs. K parameter space. The solid lines are the observed distributions and the dashed lines are the distributions predicted by our model. The model is a reasonable match, in both shape and scale, to the observed $b_J - K$ and K distributions.}
\label{ModelTest}
\end{figure}

The analysis of our sample selection supports the hypothesis that the $b_J - K$ distribution extends further and redder than observed here, as such our observed red quasar fraction is a robust lower limit of the actual red quasar fraction for $K \leq 18.4$. Weighting the quasars by the completeness of the FCSS, using the weights in Table \ref{weightingtable}, we calculated the observed red quasar fraction and found that a robust lower limit for the actual red quasar fraction is $\sim$0.22. To estimate the red quasar fraction for $K \leq 18.4$, we first assumed that the $b_J - K$ colour distribution is independent of K magnitude. Our second assumption in estimating the red quasar fraction, was that the true extent of the $b_J - K$ distribution of radio-quiet quasars should match that of radio-loud quasars; therefore, we assumed that the true extent of the $b_J - K$ distribution for our sample is $b_J - K$ = 8, matching the extent of the $b_J - K$ distribution of the PHJFS radio-loud quasars in \citet{1995Natur.375..469W}. Using these two assumptions, we postulated a model for the number density of quasars in $b_J - K$ vs. K parameter space, then used maximum likelihood estimation to find values for the model parameters; and finally integrated this model to estimate the red quasar fraction. As Figure \ref{bjMKhistogram} and the results in \citet{2003AJ....126.1131R} and \citet{2004AJ....128.1112H} are consistent with a $b_J - K$ colour distribution, where red quasars occupy a tail extending away from blue quasars, for the $b_J - K$ component of our model we used a Gamma distribution. For the K component of our model we have assumed a power-law, $10^{K\beta}$. Normalised by the number of quasars in our sample, the resultant model for the number density of quasars in $b_J - K$ vs. K parameter space is,
\begin{equation}
N(K,b_J-K) = \theta10^{K\beta}e^{-\frac{(b_J-K)}{\alpha}}(b_J-K)^{\gamma - 1}
\end{equation}
and maximum likelihood estimates of the model parameters are $\theta = 46.65$, $\beta = 0.39$, $\alpha = 0.36$ and $\gamma = 8.56$. Due to limited data, we were unable to test the validity of our model in describing the data statisitically; therefore, we assessed the model validity by plotting the observed $b_J - K$ colour and K magnitude distribution with that predicted by our model in Figure \ref{ModelTest}, and comparing them. Comparing the observed and predicted distributions in Figure \ref{ModelTest}, we concluded that the model was suitable for estimating the red quasar fraction; because it reproduced the overall shape and scale of both the K and $b_J - K$ distributions. Integrating our model over $15 \leq K \leq 18.4$, for the $b_J - K$ ranges $0 \leq b_J - K \leq 8$ and $3.5 \leq b_J - K \leq 8$, we calculated a red quasar fraction of 31\%. Using the observed $b_J - K$ distribution of our sample, we therefore estimate the $K \leq 18.4$ red quasar fraction to be 31\%, and robustly constrain it to be greater than 22\%. If the majority of red quasars are the result of dust reddening, our estimate of, and lower bound on, the red quasar fraction is consistent with that required by \citet{2007A&A...463...79G} to explain the quasar contribution to the X-Ray Background.

\section{Testing the KX Method}
\label{testKX}

\subsection{Applying the KX method}
\label{applying}
\begin{figure}
\includegraphics[width=84mm]{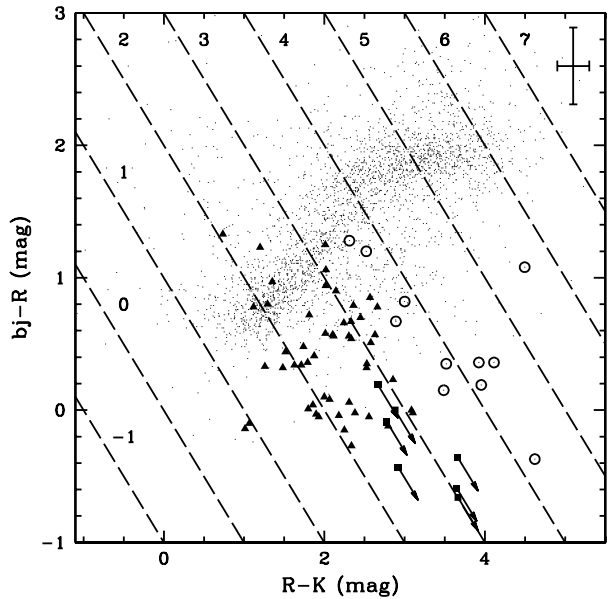}
\caption{$b_J - R$ vs. $R - K$ colour-colour plot used to test the ability of the KX method to select quasars. Dots are all stellar morphology objects in the Fornax APMCAT catalogue. Solid triangles are blue ($b_J - k <$ 3.5) quasars and circles are red ($ b_J - K \geq$ 3.5) quasars in our sample. Where an R magnitude was unavailable, limits for the $b_J - R$ and $R - K$ colours are plotted using a square and arrow. Lines of constant $b_J - K$ are plotted as dashed lines, indexed as they increase from the bottom left corner.}
\label{colourcolourplot}
\end{figure}

We tested the ability of our KX method variant to select quasars by constructing a $b_J - R$ vs. $R - K$ colour-colour plot in Figure \ref{colourcolourplot}. In Figure \ref{colourcolourplot} the majority of the quasars are distinct from the stellar locus, only 7 lie within the stellar locus, successfully demonstrating the ability of the KX method to select quasars. At the top right of Figure \ref{colourcolourplot}, the typical (mean) $b_J - R$ and $R - K$ uncertainties due to photometric uncertainties and quasar variability, are plotted as a cross. Examining the uncertainties in the quasar colours allowed us to determine how distinct from the stellar locus the quasars actually are. The photometric uncertainties in the $b_J$ and R magnitudes are both taken to be 0.1 magnitudes \citep{2000A&A...355..900D}. The K magnitude photometric uncertainties were calculated by SExtractor during processing. To account for variability, we move both the $b_J$ and K photometry to the middle epoch of the r photometry. To do this for the $b_J - R$ colour we used the structure function of \citet{1994MNRAS.268..305H}, outlined in Section \ref{detection}, with $\Delta t_{obs} =$ 15 years. We did not do this for the $R - K$ colour as a structure function was not available for the K-band, instead the variability was estimated to be 0.1 magnitudes for all the quasars, from observations in \citet{2002ApJS..141...31E}. Data from \citet{2002ApJS..141...31E} was used as they measured the K band variability of quasars satisfying $-29 < M_B < -18$ and $0 < z < 1$, which closely matches the quasar sample constructed here. For the quasars in our sample with redshifts greater than 1, 0.1 magnitude will be an over-estimate of the effect of variability; however, as it is of the order of the typical K magnitude photometric uncertainty, any over-estimate was negligible compared to the total uncertainty in the K magnitudes. Combining all the relevant uncertainties in quadrature the uncertainties in the $b_J - R$ and $R - K$ quasar colours was calculated, and typical (mean) values plotted in Figure \ref{colourcolourplot}. Examining Figure \ref{colourcolourplot}, we concluded that the quasars in our sample were indeed distinct from the stellar locus; because, the $b_J - R$ and $R - K$ uncertainties were small enough that the regions bounded by them were distinct from the stellar locus.

\subsection{Contrasting the KX method}
\label{contrast}

\begin{figure}
\includegraphics[width=84mm]{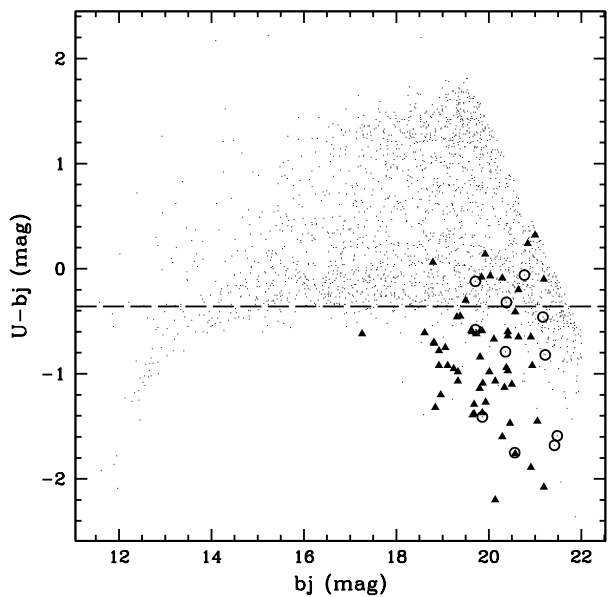}
\caption{$U - b_J$ vs $b_J$ colour-magnitude plot used to select quasars in the UVX method. Dots are all stellar morphology objects in the Fornax APM catalogue with $b_J$ and U photometry. Solid triangles are blue ($ b_J - K <$ 3.5) quasars and circles are red ($ b_J - K \geq$ 3.5) quasars in our sample with U photometry.}
\label{UVX}
\end{figure}

\begin{figure}
\includegraphics[width=84mm]{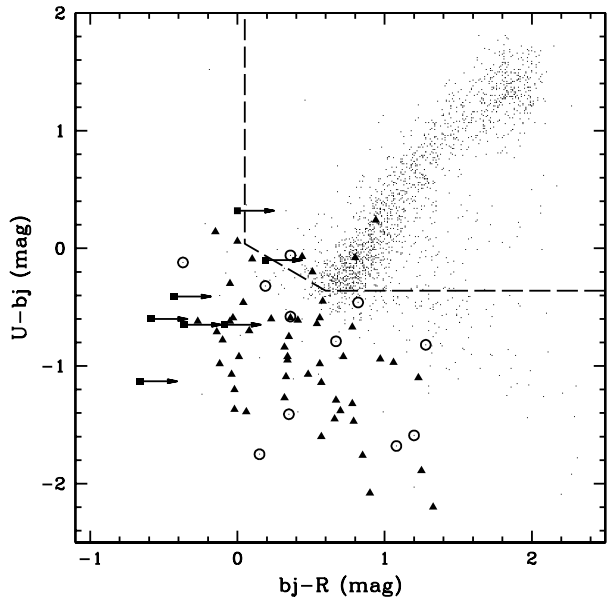}
\caption{$U - b_J$ vs $b_J - R$ colour-colour plot used to select quasars in the 2QZ survey \citep{2004mas..conf...57C}. Dots are all stellar morphology objects in the Fornax APM catalogue with $b_J$ and U photometry. Solid triangles are blue ($ b_J - K <$ 3.5) quasars and circles are red ($ b_J - K \geq$ 3.5) quasars in our sample. For the quasars in our sample without an R magnitude, squares are plotted denoting a $b_J - R$ lower limit.}
\label{multiplot}
\end{figure}

\begin{figure}
\includegraphics[width=84mm]{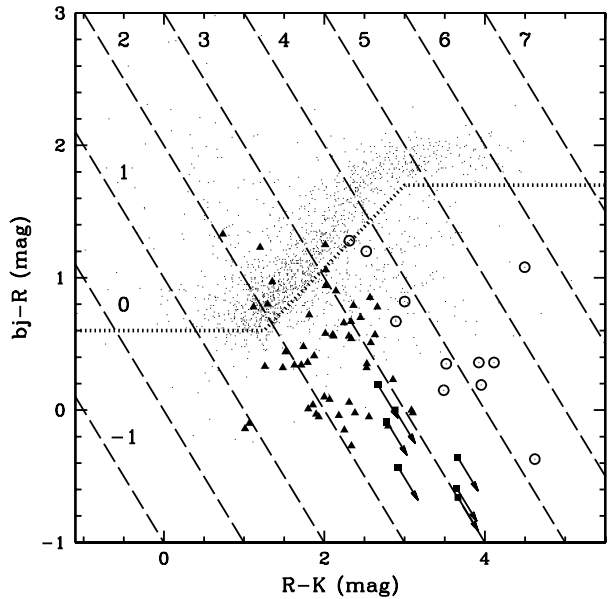}
\caption{Repeat of Figure \ref{colourcolourplot} with the added criteria that only objects with U-band photometry are plotted.}
\label{UonlyKX}
\end{figure}

\begin{table}
\caption{Example KX method criteria for selecting potential quasars for spectroscopic follow-up.}
\label{examplekxtable}
\begin{tabular}{ll}
\hline
$b_J - R$ & $R - K$ \\
\hline
($b_J - R$) $\leq 0.6$ & ($R - K$) $\leq 1.25$\\
($b_J - R$) $\leq 0.63($R - K$) - 0.19$ & $1.25 < ($R - K$) < 3$\\
($b_J - R$) $\leq 1.7$ & ($R - K$) $\geq 3$\\
\hline
\end{tabular}
\end{table}

The argument for the use of the KX method in detecting quasars is not that it finds red quasars; but that it is relatively bias free in its selection of quasars, whether red or blue. We experimentally tested this argument by comparing Figures \ref{UVX} and \ref{multiplot} to Figure \ref{UonlyKX}. Figure \ref{UVX} is a standard $U - b_J$ vs. $b_J$ colour-magnitude plot used in selecting quasars in the UVX approach. Figure \ref{multiplot} is a $U - b_J$ vs $b_J - R$ colour-colour plot used in the 2QZ to select quasars \citep{2004mas..conf...57C}. As Figures \ref{UVX} and \ref{multiplot} only contain objects that have U-band photometry, in Figure \ref{UonlyKX} we generated the same colour-colour plot in Figure \ref{colourcolourplot}, but only for objects that had U-band photometry so as to make a fair comparison. Comparing Figures \ref{UVX} and \ref{multiplot} to Figure \ref{UonlyKX}, qualitatively, the KX method immediately appeared more effective than the UVX method and comparable to the 2QZ multicolour method. We examined this quantitatively by measuring the percentage of quasars in our sample that would not be selected by the UVX and 2QZ multi-color methods, in total and as a function of colour, and compared this to our KX method variant. In Figure \ref{UVX} we use a colour criterion of $U - b_J <$ -0.36 to select quasars as an example of a typical UVX method criterion, based on Smith et al. (1997), \citet{1990MNRAS.243....1B,1997AJ....113.1517L}. Using this cut 13 of the 69 quasars, or 19\%, were not selected in Figure \ref{UVX}. Breaking this into red and blue quasars, where red quasars satisfy $ b_J - K \geq$ 3.5, 3 (10) of 11 (58) red (blue) quasars were not selected. The UVX method failed to select 27\% of the red quasars compared to 17\% of the blue quasars. In Figure \ref{multiplot} we applied the selection criterion used in the 2QZ \citep{2004mas..conf...57C}, the criterion is marked by a dashed line and objects below or to the left of the line were selected as quasars. This selection criterion failed to select 5 of the 69 quasars, corresponding to 7\% of the quasars. Broken into red and blue quasars, 1 red and 4 blue quasars were missed, equating to 9\% and 7\%. For the KX method variant used here, in Figure \ref{UonlyKX} we applied an example of a suitable colour-based selection criterion, and plotted it as a dotted line in Figure \ref{UonlyKX} with objects below the line selected as quasars. The details of our example selection criterion are listed in Table \ref{examplekxtable}. Our example selection criterion fails to select 7 quasars, 1 red and 6 blue, corresponding to 10\% of all the quasars and 9\% and 10\% of the red and blue quasars respectively. As we knew where the quasars were located on the colour-colour plot before defining a suitable colour-based selection criterion; comparing the results of our example KX method selection criterion to the results of the UVX and 2QZ methods is potentially an unfair comparison. As an alternative, we used the proximity to the stellar locus as a selection criterion i.e. whether a quasar was obscured by the stellar locus. In Figure \ref{UonlyKX}, 7 quasars were clearly obscured by the stellar locus, 1 red and 6 blue. This corresponds to 10\% of all the quasars and 9\% of both the red and blue quasars. Comparing all these percentages we found that the KX method variant was quantitatively superior to the UVX method, as the UVX method selected less quasars and was clearly biassed towards the selection of blue quasars. We also found that quantitatively the KX method variant is comparable to the 2QZ multicolour method in the number of quasars selected; however, the KX method variant is superior as the 2QZ multicolour did demonstrate a bias towards the selection of blue quasars while the KX method variant did not. Our results are strong experimental support that compared to existing methods such as the UVX and 2QZ multicolor methods, the KX method, and in particular our variant, is ideal for selecting quasar candidates for spectroscopic follow-up, independent of quasar colour. The key to this result is that the lines of constant $b_J - K$ in Figure \ref{UonlyKX} cut through the stellar locus almost perpendicularly; therefore, quasar $b_J - K$ colour does not affect the likelihood of quasar selection using our KX method variant.

\section{Conclusions}
\label{conclusion}
In this paper, we constructed a sample of 69 quasars with measured $b_J - K$ colours, using a combination of APMCAT, IRIS2 imaging and FCSS spectroscopic identifications. 11 of these quasars are red, satisfying $ b_J - K \geq$ 3.5, and all of these red quasars are radio-quiet according to the All Sky Optical Catalogue of Radio/X-Ray Sources. In accordance with \citet{2001AJ....121.2843B} and \citet{2006MNRAS.367..717M}, as the 11 red quasars found here are unresolved, luminous and radio-quiet sources predominantly located at $z > 1$, this strongly indicates that the main cause of the red $b_J - K$ colours is dust at the quasar redshift. Comparing our quasar sample to LBQS quasars, we found that the two $b_J - K$ colour distributions are different at the 99.9\% confidence level, with our sample having a significantly broader, redder distribution. Analysis of the uncertainties in the $b_J - K$ colours demonstrated that neither our red quasar detections or our comparison to the LBQS quasars is affected by them. A second analysis, on the effects of our datasets magnitude and completeness limits, revealed that they limited our measured $b_J - K$ distribution to $ b_J - K <$ 5. As red quasars are observed up to this colour limit, we concluded that the true $b_J - K$ distribution of quasars is even broader and redder than observed here. From the observed $b_J - K$ colour distribution, we robustly constrained the red quasar fraction of the $K \leq 18.4$ quasar population to be greater than 22\%, and from a model estimated it to be 31\%.

Using the quasar sample constructed here, the viability of a KX method variant was tested. Using a $b_J - R$ vs $R - K$ plot, the KX method variant was capable of separating the quasar sample out from the other objects in the APMCAT catalogue. Comparing the KX method variant to the UVX and 2QZ multicolour methods, the KX method variant was found to be as effective in the number of quasars it selected; however, it was superior at selecting quasars independent of colour. For those reasons, the KX method variant used here is an ideal technique for future large surveys to use in selecting potential quasars, whether red or blue, for spectroscopic follow-up.

\section*{Acknowledgments}
\label{acknowledgements}
We thank the AAO for use of the IRIS2 instrument and Stuart Ryder of the AAO for his help in reducing the IRIS2 imaging.

\label{lastpage}

\end{document}